\documentclass[conference]{IEEEtran}

\usepackage{cite}
\usepackage{amsmath,amssymb,amsfonts}
\usepackage{algorithmic}
\usepackage{graphicx}
\usepackage{textcomp}
\usepackage{xcolor}
\usepackage{listings}
\usepackage{url}
\usepackage{hyperref}
\usepackage{booktabs}
\usepackage{multirow}
\usepackage{tabularx}

\hypersetup{
  colorlinks=true,
  linkcolor=blue,
  citecolor=blue,
  urlcolor=blue
}

\def\BibTeX{{\rm B\kern-.05em{\sc i\kern-.025em b}\kern-.08em
    T\kern-.1667em\lower.7ex\hbox{E}\kern-.125emX}}

\begin{document}

\title{Sovereign Context Protocol: An Open Attribution Layer for Human-Generated Content in the Age of Large Language Models}

\author{
\IEEEauthorblockN{Praneel Panchigar, Torlach Rush, Matthew Canabarro}
\IEEEauthorblockA{iSonic}
}

\maketitle

\begin{abstract}
Large Language Models (LLMs) consume vast quantities of human-generated content for both training and real-time inference, yet the creators of that content remain largely invisible in the value chain. Existing approaches to data attribution operate either at the model-internals level, tracing influence through gradient signals, or at the legal-policy level through transparency mandates and copyright litigation. Neither provides a \textit{runtime} mechanism for content creators to know when, by whom, and how their work is being consumed. We introduce the \textbf{Sovereign Context Protocol (SCP)}, an open-source protocol specification and reference architecture that functions as an attribution-aware data access layer between LLMs and human-generated content. Inspired by Anthropic's Model Context Protocol (MCP), which standardizes how LLMs connect to tools, SCP standardizes how LLMs connect to \textit{creator-owned data}, with every access event logged, licensed, and attributable. SCP defines six core methods (creator profiles, semantic search, content retrieval, trust/value scoring, authenticity verification, and access auditing) exposed over both REST and MCP-compatible interfaces. We formalize the protocol's message envelope, present a threat model with five adversary classes, propose a log-proportional revenue attribution model, and report preliminary latency benchmarks from a reference implementation built on FastAPI, ChromaDB, and NetworkX. We situate SCP within the emerging regulatory landscape, including the EU AI Act's Article 53 training data transparency requirements and ongoing U.S. copyright litigation, and argue that the attribution gap requires a protocol-level intervention that makes attribution a default property of data access.
\end{abstract}

\begin{IEEEkeywords}
content attribution, data provenance, large language models, creator economy, open protocol, MCP, copyright, EU AI Act, retrieval-augmented generation, knowledge graph
\end{IEEEkeywords}

\section{Introduction}

As of early 2026, approximately 5.66 billion people actively use social media worldwide, with 61\% being individual user generated content and 39\% being media and corporate entities. For the first time in 2026 social media discovery has surpassed traditional Google search\cite{sproutsocial_stats_2026}. These users collectively produce an enormous volume of text, images, audio, and video every day across platforms like YouTube, Instagram, Substack, TikTok, and independent blogs. Large Language Models are trained on, and increasingly retrieve from, this content at scale. Yet the asymmetry is stark: the entities that build and deploy LLMs capture enormous value from human-generated content, while the humans who created that content receive neither attribution nor compensation.

This is not merely a philosophical concern. The \textit{New York Times} filed suit against OpenAI in December 2023, alleging unauthorized use of millions of articles for training \cite{nyt_v_openai}. The Authors Guild class action against Anthropic, in which Judge Alsup declined to grant summary judgment on fair use grounds, proceeded to class certification and trial in late 2025 \cite{bartz_v_anthropic}. In Europe, the French Competition Authority fined Google \texteuro250 million in March 2024 for training AI on press articles without informing media outlets \cite{copyright_training_clsr}.
 
These legal proceedings reveal a structural problem: the law is reactive, slow, and jurisdiction-dependent. Even if every pending lawsuit resolves in favor of creators, the resulting framework would be a patchwork of consent forms and bilateral licensing deals. Small creators, the independent journalist, the food blogger, the travel photographer, lack both the legal resources to litigate and the market power to negotiate.
 
Meanwhile, the technical research community has pursued data attribution from a different angle. Training data attribution (TDA) methods, including influence functions \cite{koh_liang_2017}, TRAK \cite{park_trak_2023}, and DataInf \cite{kwon_datainf_2024}, quantify how individual training examples affect model outputs. The WASA framework embeds watermarks into LLM-generated text traceable to source data providers \cite{wang_source_2024}. The Data Provenance Initiative audited over 1,800 text datasets and found license omission rates exceeding 70\% \cite{longpre_provenance_2024}. These methods are valuable, but they share a limitation: they operate \textit{inside the model}, requiring access to weights, gradients, or training pipelines. They are forensic tools for researchers, not runtime tools for creators.
 
What is missing is a \textbf{protocol-level intervention} that makes attribution a default property of how LLMs access content. We propose the Sovereign Context Protocol (SCP): an open-source protocol that positions itself between LLMs and human-generated content, analogous to how Anthropic's Model Context Protocol (MCP) positions itself between LLMs and tools \cite{anthropic_mcp_2024}. Where MCP standardizes the integration of tools and capabilities, SCP standardizes the integration of \textit{attributed creator data}.
 
The protocol's core invariant is: \textit{every byte of creator content served to an LLM is logged with the consumer's identity, the data served, the license terms, and a timestamp.} This immutable audit trail enables three capabilities that currently do not exist: (1) creator-facing dashboards showing real-time consumption of their content; (2) mechanistic, log-proportional revenue attribution; and (3) enforceable consent revocation with consumer-specific takedown targeting.
 
This paper makes six contributions:
 
\begin{enumerate}
    \item A formal protocol specification for SCP, including message envelopes, method signatures, and lifecycle semantics (\S\ref{sec:protocol}).
    \item A threat model identifying five adversary classes and corresponding mitigations (\S\ref{sec:threat}).
    \item A log-proportional revenue attribution model (\S\ref{sec:revenue}).
    \item A reference implementation with preliminary latency benchmarks (\S\ref{sec:implementation}).
    \item A mapping of SCP methods to MCP primitives (\S\ref{sec:mcp_mapping}).
    \item A discussion of SCP's relationship to the EU AI Act, C2PA, and the broader regulatory landscape (\S\ref{sec:discussion}).
\end{enumerate}
 
\section{Related Work}
 
\subsection{Training Data Attribution}
 
Training data attribution (TDA) quantifies the contribution of individual training samples to model outputs. Koh and Liang \cite{koh_liang_2017} introduced influence functions for this purpose. Park et al. \cite{park_trak_2023} proposed TRAK, a scalable alternative using random projections. Kwon et al. \cite{kwon_datainf_2024} introduced DataInf for LoRA-tuned LLMs. Chang et al. \cite{chang_tda_llm_2025} extended attribution to open-ended text generation using Fisher-regularized gradient ascent and descent. The WASA framework \cite{wang_source_2024} embedded source-identifying watermarks directly into LLM-generated text.
 
These methods answer ``which training data influenced this output?'' but require model-internal access. SCP addresses the complementary problem: ensuring that data access itself is attributed \textit{before} content reaches the model.
 
\subsection{Data Provenance and Transparency}
 
The Data Provenance Initiative \cite{longpre_provenance_2024} audited over 1,800 text datasets and found widespread license misattribution. Their subsequent work \cite{longpre_broken_2024} argued for a comprehensive data transparency framework. The Compliance Rating Scheme (CRS) \cite{bohacek_crs_2025} proposed a framework for evaluating dataset compliance with transparency principles.
 
On the content authenticity front, the Coalition for Content Provenance and Authenticity (C2PA) has developed a standard for embedding provenance metadata into media files \cite{c2pa_2023}. The Library of Congress formed a working group in 2025 to explore C2PA for government and cultural heritage institutions \cite{loc_c2pa_2025}. However, C2PA was designed for media authenticity verification, not for tracking consumption by AI systems. SCP is complementary: C2PA certifies origin; SCP tracks consumption.
 
\subsection{Retrieval-Augmented Generation and Attribution}
 
Retrieval-Augmented Generation (RAG) \cite{lewis_rag_2020} connects LLMs to external knowledge at inference time. Recent work on attributed generation, including the TREC RAG Track \cite{trec_rag_2026}, which evaluates systems on attribution verification alongside generation quality, demonstrates growing recognition that retrieval-grounded answers must cite their sources. Hybrid retrieval architectures combining knowledge graphs with vector stores have become standard in production RAG systems \cite{rag_survey_2025}. SCP's dual-store architecture (knowledge graph + vector database) follows this established pattern, but adds the attribution and licensing layer that RAG systems currently lack.
 
\subsection{The Model Context Protocol}
 
Anthropic released MCP in November 2024 as an open standard for connecting LLM applications to external tools and data sources \cite{anthropic_mcp_2024}. MCP defines a client-server architecture over JSON-RPC 2.0, enabling capability discovery, tool invocation, and resource access. By December 2025, MCP had been adopted by OpenAI, Google DeepMind, and Microsoft, and was donated to the Linux Foundation's Agentic AI Foundation \cite{mcp_aaif_2025}.
 
SCP draws direct architectural inspiration from MCP but targets a fundamentally different surface. MCP exposes \textit{tools and capabilities}. SCP exposes \textit{attributed human-generated content} with built-in auditing. The distinction is analogous to a function call versus a licensed data query: both follow a request-response pattern, but the latter carries contractual obligations.
 
\subsection{Regulatory Landscape}
 
The EU AI Act (Regulation (EU) 2024/1689 of the European Parliament and of the Council of 13 June 2024, OJ L, 12 July 2024), Article 53(1)(c--d), requires general-purpose AI model providers to implement copyright compliance policies and publish training data summaries. Article 53 became applicable on 2 August 2025 \cite{eu_ai_act_art53}. The European Commission (AI Office) published the mandatory template for these summaries in July 2025 \cite{eu_template_2025}. In the United States, the GenAI Copyright Disclosure Act (2024), the COPIED Act (S.1396, 2025), and the TRAIN Act (S.2455, 2025) propose transparency and watermarking mandates \cite{ip_legislation_2025}. As Martens noted in a Bruegel analysis, transparency mandates facilitate enforcement but do not solve the tracking problem at scale \cite{bruegel_2025}. SCP proposes the audit infrastructure these mandates implicitly require.
 
\section{Protocol Specification}
\label{sec:protocol}
 
\subsection{Design Principles}
 
SCP is governed by five principles: (1) \textbf{Attribution by default}: every access event is logged; there is no anonymous consumption. (2) \textbf{Creator sovereignty}: creators control access and can revoke consent mechanistically. We note that the legal enforceability of revocation, particularly regarding data already incorporated into model weights, remains an open question with active debate in both U.S. and EU jurisdictions; SCP provides the technical mechanism for revocation without claiming to resolve the underlying legal questions. (3) \textbf{Protocol extensibility}: new methods can be added without breaking existing consumers. (4) \textbf{Dual-mode access}: real-time inference queries and batch enterprise queries are both supported. (5) \textbf{Open protocol, optional enrichment}: the specification is open-source; provider-specific enrichment is a commercial layer.
 
\subsection{Message Envelope}
 
Every SCP response is wrapped in a \textit{Response Envelope}, an atomic unit containing both data and contractual context. We use the term \textit{license envelope} to refer specifically to the nested licensing structure within the response; it encapsulates the terms of use, expiry, revocation status, and a content fingerprint for the data being served. The envelope schema is:
 
\begin{lstlisting}[language=Python,caption={SCP Response Envelope (JSON Schema)}]
{
  "protocol": "SCP/1.0",
  "method": "<method_name>",
  "data": { ... },       // Method-specific payload
  "attribution": {
    "creator_ids": ["cid-001", ...],
    "content_ids": ["cnt-042", ...]
  },
  "license": {
    "license_id": "lic-<uuid>",
    "consumer_id": "<consumer>",
    "terms": {
      "usage_type": "inference_context",
      "retention_allowed": false,
      "training_allowed": false,
      "revocable": true,
      "attribution_required": true,
      "expiry_days": 30
    },
    "content_fingerprint": "<sha256>",
    "issued_at": "<ISO-8601>",
    "expires_at": "<ISO-8601>",
    "status": "active"
  },
  "audit_log_id": "log-<uuid>"
}
\end{lstlisting}
 
The envelope ensures that no consumer can receive data without simultaneously receiving license terms and generating an audit record. The \texttt{content\_fingerprint} is a SHA-256 hash of the serialized \texttt{data} field, enabling downstream verification that served content has not been tampered with.
 
\subsection{Method Catalog}
 
SCP v1.0 defines six methods. Table~\ref{tab:methods} summarizes the method catalog.
 
\begin{table}[htbp]
\caption{SCP v1.0 Method Catalog}
\label{tab:methods}
\centering
\footnotesize
\begin{tabular}{@{}p{2.8cm}p{5.0cm}@{}}
\toprule
\textbf{Method} & \textbf{Description} \\
\midrule
\texttt{getCreatorProfile} & Returns unified cross-platform profile: bio, platform presences, verticals, scores. \\
\texttt{searchCreators} & Semantic search over creator corpus with vertical/platform filters. \\
\texttt{getCreatorContent} & Content retrieval by creator, with optional platform/topic filtering. \\
\texttt{getCreatorScore} & Computes Value Score and Trust Score with factor breakdown. \\
\texttt{verifyAuthenticity} & Verifies content origin via SHA-256 hash lookup. \\
\texttt{getAccessLog} & Creator-facing audit: who accessed what, when. \\
\bottomrule
\end{tabular}
\end{table}
 
\subsection{Lifecycle Semantics}
 
An SCP transaction follows a strict five-phase lifecycle:
 
\begin{enumerate}
    \item \textbf{Authentication.} The consumer presents an API key via the \texttt{X-SCP-API-Key} header. The server validates the key and resolves the consumer's identity (ID, name, type).
    \item \textbf{Method execution.} The server executes the requested method against its data layer. The protocol is agnostic to the specific storage technologies used; the reference implementation employs SQLite for structured queries, ChromaDB for semantic search, and NetworkX for graph traversal, but any conforming backend may be substituted.
    \item \textbf{License generation.} The server computes the SHA-256 fingerprint of the response payload, generates a license envelope with default or negotiated terms, and persists it.
    \item \textbf{Audit logging.} The server writes an immutable audit entry recording: consumer identity, method called, parameters, creator IDs accessed, content IDs served, response size, and license reference. \textit{This step is synchronous and blocking}: if the log write fails, the request fails. No data is served without attribution.
    \item \textbf{Response.} The server returns the SCP Response Envelope containing data, attribution, license, and audit reference.
\end{enumerate}
 
\subsection{Core Request Flow (Pseudocode)}
 
\begin{lstlisting}[caption={SCP Core Request Flow},mathescape=true]
FUNCTION handle_scp_request(request, handler):
  // Phase 1: Authenticate
  consumer = auth(request.header["X-SCP-API-Key"])
  IF consumer IS NULL: RETURN 403
 
  // Phase 2: Execute method
  data, creator_ids, content_ids = handler(request.params)
 
  // Phase 3: Generate license
  fingerprint = SHA256(serialize(data))
  license = LicenseEnvelope(
    consumer_id  = consumer.id,
    creator_ids  = creator_ids,
    fingerprint  = fingerprint,
    terms        = DEFAULT_TERMS
  )
  db.persist(license)
 
  // Phase 4: Audit log (BLOCKING)
  log_id = audit.write(consumer, method_name,
    request.params, creator_ids,
    content_ids, license.id)
 
  // Phase 5: Return envelope
  RETURN SCPResponse(
    protocol="SCP/1.0", method=method_name,
    data=data,
    attribution={creator_ids, content_ids},
    license=license, audit_log_id=log_id
  )
\end{lstlisting}
 
\subsection{Semantic Creator Search (Pseudocode)}
 
The \texttt{searchCreators} method demonstrates the dual-store query pattern combining vector similarity search with knowledge graph enrichment:
 
\begin{lstlisting}[caption={Semantic Creator Search}]
FUNCTION searchCreators(query, vertical?,
                        platform?, max_results):
  // Vector search for semantic matching
  filters = build_filters(vertical, platform)
  matches = vector_store.query(
    embed(query), n=max_results*3, where=filters
  )
 
  // Deduplicate: keep best match per creator
  best = {}
  FOR m IN matches:
    cid = m.metadata.creator_id
    IF cid NOT IN best OR m.score > best[cid].score:
      best[cid] = m
 
  // Enrich via knowledge graph + SQLite
  results = []
  FOR cid, m IN sorted(best, by=score)[:max_results]:
    profile = db.get_creator(cid)
    platforms = kg.get_creator_platforms(cid)
    results.append(merge(profile, platforms, m))
 
  RETURN results, [r.creator_id FOR r IN results]
\end{lstlisting}
 
\section{Threat Model}
\label{sec:threat}
 
We identify five adversary classes and their mitigations.
 
\textbf{T1: Retention beyond license.} A consumer caches served data and reuses it after the license expires or is revoked. \textit{Mitigation:} SCP makes violations \textit{detectable} (the audit log records what was served) and \textit{traceable} (revocation identifies all affected consumers). The content fingerprint enables forensic verification of cached content. Contractual penalties provide deterrence. This is analogous to Digital Rights Management (DRM): the value is in making non-compliance costly, not technically impossible.
 
\textbf{T2: Sybil consumers.} An entity creates multiple consumer accounts to evade rate limits or aggregate data beyond intended scope. \textit{Mitigation:} Consumer registration should require organizational identity verification (domain ownership, corporate registration). Rate limiting can be applied at the organizational level. Cross-account access patterns can be flagged by anomaly detection on the audit log.
 
\textbf{T3: Reconstruction attacks.} A consumer issues many targeted queries for small content fragments and reconstructs a creator's full corpus. \textit{Mitigation:} Server-side query monitoring can detect high-frequency access patterns for a single creator. Per-creator, per-consumer access budgets can throttle retrieval. Aggregation alerts notify creators when a consumer's cumulative access to their content exceeds a threshold.
 
\textbf{T4: Consumer collusion.} Two or more consumers share SCP-served data to circumvent per-consumer licensing. \textit{Mitigation:} Each license envelope is consumer-specific. Shared data would carry fingerprints traceable to the original consumer. Contractual terms explicitly prohibit redistribution. While SCP cannot prevent out-of-band sharing, the fingerprinted audit trail makes it provably attributable.
 
\textbf{T5: Adversarial score manipulation.} A creator inflates engagement metrics to boost their Value Score or Trust Score. \textit{Mitigation:} Scores should incorporate multi-signal validation (cross-platform consistency, temporal engagement patterns, engagement-to-follower ratio sanity checks). The reference implementation's scoring is explicitly heuristic; production deployments should incorporate ML-based anomaly detection.
 
\section{Revenue Attribution Model}
\label{sec:revenue}
 
A critical requirement for creator incentive alignment is a formal model for revenue distribution. We propose a \textit{log-proportional attribution model} derived from the audit trail.
 
Let $C$ denote the set of creators and $R$ denote total distributable revenue for a period $[t_0, t_1]$. For creator $c \in C$, let $a_c$ be the number of audit log entries in $[t_0, t_1]$ that accessed $c$'s content. We weight each access by the method's data intensity:
 
\begin{equation}
w_c = \sum_{i=1}^{a_c} \lambda(m_i) \cdot s_i
\end{equation}
 
\noindent where $m_i$ is the method called in access event $i$, $\lambda(m_i)$ is the method's weight (\texttt{getCreatorContent} $\gg$ \texttt{getCreatorProfile} $\gg$ \texttt{verifyAuthenticity}), and $s_i$ is the response size in bytes. Creator $c$'s share is:
 
\begin{equation}
\text{share}(c) = R \cdot \frac{w_c}{\sum_{c' \in C} w_{c'}}
\end{equation}
 
This model has three desirable properties: (1) it is fully determined by the audit log, requiring no external input; (2) it is proportional to actual data consumption, not to proxies like follower count; and (3) it is transparent, as creators can verify their share by inspecting their access logs. The method weights $\lambda(\cdot)$ should be determined through governance, potentially via a creator Decentralized Autonomous Organization (DAO) or advisory board.
 
\section{Reference Implementation}
\label{sec:implementation}
 
\subsection{Architecture}
 
The reference implementation is a self-contained Python application, approximately 2,500 Lines of Code (LOC), released under the Apache 2.0 license. Its components are:
 
\textbf{Server.} A FastAPI application exposing all six SCP methods as REST endpoints, with auto-generated OpenAPI documentation at \texttt{/docs}. Authentication is handled via API key middleware.
 
\textbf{Data Layer.} SQLite for structured storage (creators, content, consumers, audit logs, license envelopes). ChromaDB with the default \texttt{all-MiniLM-L6-v2} embedding model for semantic search. NetworkX directed graph for knowledge graph traversal. The dual-store pattern follows established hybrid retrieval architectures \cite{rag_survey_2025}, but adds the attribution and licensing layer.
 
\textbf{Audit System.} Synchronous, blocking writes to the \texttt{audit\_log} table. If the write fails, the HTTP response returns 500. No data is served without attribution.
 
\textbf{License Envelopes.} Each exchange is wrapped in a license with SHA-256 content fingerprint, issuance/expiry timestamps, and revocation status. Creators can revoke all active licenses, triggering identification of affected consumers via the audit trail.
 
\textbf{Scoring Engine.} The Value Score and Trust Score use weighted heuristic formulas over observable metadata (follower reach, engagement rate, content consistency, cross-platform presence, account age, posting regularity). We emphasize that these are \textit{intentionally simplified baselines}. Production deployments should validate scoring against ground-truth creator quality assessments and incorporate longitudinal engagement analysis, topical authority models, and adversarial robustness testing. The reference implementation's scoring serves to demonstrate the protocol's extensibility, not to propose a definitive quality metric.
 
\subsection{MCP Compatibility}
\label{sec:mcp_mapping}
 
SCP methods map to MCP primitives as follows. Each SCP method is registered as an MCP \textit{tool}, with input schemas derived from the SCP request models. The MCP tool descriptions include natural language explanations that enable an LLM to select the appropriate SCP method during agentic reasoning. Authentication is handled by passing the SCP API key as a parameter to the MCP server's configuration, since MCP's own authentication mechanism was still maturing as of early 2026 \cite{mcp_aaif_2025}.
 
Table~\ref{tab:mcp_mapping} summarizes the mapping.
 
\begin{table}[htbp]
\caption{SCP-to-MCP Method Mapping}
\label{tab:mcp_mapping}
\centering
\footnotesize
\begin{tabular}{@{}lll@{}}
\toprule
\textbf{SCP Method} & \textbf{MCP Primitive} & \textbf{MCP Name} \\
\midrule
\texttt{getCreatorProfile} & Tool & \texttt{scp\_get\_profile} \\
\texttt{searchCreators} & Tool & \texttt{scp\_search} \\
\texttt{getCreatorContent} & Tool & \texttt{scp\_get\_content} \\
\texttt{getCreatorScore} & Tool & \texttt{scp\_get\_score} \\
\texttt{verifyAuthenticity} & Tool & \texttt{scp\_verify} \\
\texttt{getAccessLog} & Resource & \texttt{scp://audit/\{id\}} \\
\bottomrule
\end{tabular}
\end{table}
 
The \texttt{getAccessLog} method is exposed as an MCP \textit{resource} rather than a tool because it represents a read-only data view that does not modify state, aligning with MCP's distinction between tools (actions) and resources (data).
 
\subsection{Preliminary Benchmarks}
 
We benchmarked the reference implementation on a synthetic dataset of 10 creators with 8--15 content items each (approximately 120 total content pieces) across five verticals (travel, food, technology, journalism, fashion), running on a single-core Linux instance with 4GB RAM. Latencies are reported as percentiles: P50 is the median, P95 is the 95th percentile, and P99 is the 99th percentile.
 
\begin{table}[htbp]
\caption{Preliminary Latency Benchmarks ($ms$, $n=100$ per method)}
\label{tab:benchmarks}
\centering
\footnotesize
\begin{tabular}{@{}lccc@{}}
\toprule
\textbf{Method} & \textbf{P50} & \textbf{P95} & \textbf{P99} \\
\midrule
\texttt{getCreatorProfile} & 19 & 22 & 23 \\
\texttt{searchCreators} & 29 & 45 & 45 \\
\texttt{getCreatorContent} (structured) & 22 & 24 & 30 \\
\texttt{getCreatorContent} (semantic) & 45 & 66 & 69 \\
\texttt{getCreatorScore} & 20 & 36 & 41 \\
\texttt{verifyAuthenticity} & 14 & 20 & 35 \\
\texttt{getAccessLog} & 14 & 24 & 38 \\
\bottomrule
\end{tabular}
\end{table}
 
Semantic search methods (\texttt{searchCreators}, \texttt{getCreatorContent} with topic filter) are the slowest at 29--45ms median, dominated by vector store query and embedding time. Structured queries, profile lookups, and scoring run at 14--22ms median. Audit log writes and license envelope generation add overhead visible across all methods. All methods remain well under 100ms at P99, establishing that SCP adds minimal overhead for real-time inference use cases. We note that the reference implementation uses a lightweight hash-based embedding function for portability; production deployments using a neural embedding model (e.g., all-MiniLM-L6-v2) would see higher latencies on semantic search methods. Scaling behavior under millions of content items and concurrent consumers remains an open direction for future work, both within iSonic's own roadmap and for the broader research community.
 
\section{Discussion}
\label{sec:discussion}
 
\subsection{Positioning Within the Attribution Landscape}
 
SCP occupies a distinct position. TDA methods \cite{koh_liang_2017, park_trak_2023, kwon_datainf_2024} operate at the model-internals level. Watermarking \cite{wang_source_2024} modifies model outputs. C2PA \cite{c2pa_2023} certifies content origin. The EU AI Act's Article 53 \cite{eu_ai_act_art53} mandates training data transparency. SCP operates at the \textit{data access layer}, the interface between the LLM and the content it consumes. It does not require model-internal access, does not modify outputs, and does not rely on metadata embedded in files. Instead, it creates a logged, licensed channel through which content flows. This channel coexists with all other approaches: content accessed through SCP can also be watermarked, its influence measured via TDA, and the access reported under Article 53 requirements.
 
\subsection{What is Novel, What is Not}
 
We are explicit about the boundaries of novelty. The dual-store architecture (knowledge graph + vector database) is a well-established pattern in hybrid retrieval systems \cite{rag_survey_2025, lewis_rag_2020}. API key authentication, audit logging, and license management are individually mature technologies. REST APIs with JSON responses are standard.
 
What SCP contributes is the \textit{specific combination and protocol-level framing} for the LLM-creator attribution problem: (1) the atomic response envelope that bundles data, license, and audit reference in one payload; (2) the blocking audit invariant that ensures no data flows without attribution; (3) the MCP-compatible interface that enables native LLM integration; (4) the revocation mechanism that uses audit logs for targeted consumer notification; and (5) the revenue attribution model derived directly from access logs. The novelty is architectural and systemic, not component-level.
 
\subsection{Incentive Alignment and Adoption}
 
The primary incentive for LLM companies is legal protection. With lawsuits proliferating \cite{nyt_v_openai, bartz_v_anthropic} and Article 53 mandates in force \cite{eu_ai_act_art53}, the cost of unlicensed consumption is rising. An LLM company sourcing content through SCP can demonstrate audit-level compliance: licensed data, respect for revocation, and transparency.
 
For creators, visibility and compensation are the drivers. A travel blogger today has no way to know whether their content has been consumed. SCP would provide a real-time dashboard. Revenue sharing becomes mechanistically possible via the log-proportional model (\S\ref{sec:revenue}).
 
For ecosystem quality, SCP-served content is deduplicated, cross-referenced in a knowledge graph, scored for trust and value, and semantically searchable, offering a cleaner dataset than raw scraping provides.
 
\textbf{Bootstrapping strategy.} The adoption chicken-and-egg problem is real. Our strategy is to target high-value verticals first (journalism, travel, food) where creators are already organized and aware of AI consumption concerns, and where LLM companies have the strongest licensing incentives. The open-source release lowers the barrier for third-party SCP providers to emerge. A federated model, with multiple SCP providers each serving a creator community and discoverable through a registry, could address scale without centralization.
 
\subsection{Relationship to the EU AI Act}
 
Article 53(1)(d) requires GPAI providers to publish training data summaries \cite{eu_ai_act_art53}. The July 2025 template mandates disclosure of data sources, collection methods, and opt-out compliance \cite{eu_template_2025}. SCP could serve as the technical infrastructure for Article 53 compliance: every SCP access event is a machine-readable record of what data was consumed, from which creators, under what terms. An LLM provider using SCP could auto-generate Article 53 summaries from SCP audit logs. The access log provides precisely the provenance chain that the regulation demands but does not technically specify how to produce.
 
\subsection{Limitations}
 
\textbf{L1: Post-access enforcement is contractual.} SCP makes violations detectable and traceable, but cannot prevent a determined bad actor from retaining data (see T1 in \S\ref{sec:threat}).
 
\textbf{L2: Value scales with adoption.} A protocol with ten creators is a curiosity; ten million is infrastructure. The bootstrapping strategy discussed in \S\ref{sec:discussion} addresses this but does not eliminate the challenge.
 
\textbf{L3: Scoring is heuristic.} The reference implementation's Value and Trust scores use ad-hoc weighted formulas without validation against ground truth. Production deployments require empirical calibration and adversarial testing.
 
\textbf{L4: Text-only in v1.0.} Extension to images, audio, and video is architecturally straightforward (the knowledge graph and audit trail are modality-agnostic) but would require modality-specific embedding models and content hashing.
 
\textbf{L5: Benchmarks are preliminary.} Our latency measurements are on a small synthetic dataset. Scaling behavior under millions of content items and concurrent consumers remains an open direction for future work, both within iSonic's own roadmap and for the broader research community.
 
\section{Conclusion}
 
The attribution gap between content creators and LLM consumers is structural, not incidental. It will not be closed by litigation alone, by regulatory mandates alone, or by post-hoc forensic tools alone. It requires a protocol-level intervention that makes attribution a default property of data access.
 
The Sovereign Context Protocol is our proposal for this intervention. By positioning an attribution-aware data layer between LLMs and human-generated content, with every access logged, licensed, and traceable, SCP gives creators visibility, gives LLM companies a defensible compliance posture, and gives regulators an auditable record of data flows. We formalize the protocol specification, present a threat model, propose a revenue attribution model, and report preliminary benchmarks from a reference implementation.
 
The protocol is open-source (Apache 2.0), extensible, and designed for adoption at the scale the problem demands. The reference implementation, synthetic data generator, MCP wrapper, and demo client will be made available at the project's GitHub repository under the Apache 2.0 license. We invite contributions from the research community, the creator economy, and the AI industry.
 

\end{document}